\documentclass{webofc}

\usepackage[varg]{txfonts}
\usepackage[dvipsnames]{xcolor}
\usepackage{subfigure}
\usepackage{IEEEtrantools}
\usepackage{hyperref}

\DeclareMathOperator{\Tr}{Tr}
\newcommand{\A}{\mathrm{A}}
\newcommand{\B}{\mathrm{B}}
\newcommand{\p}{\partial}
\newcommand{\ev}[1]{\big<  #1 \big> }
\newcommand{\vv}[1]{\mathbf{#1}}
\newcommand{\dg}{\dagger}

\newcommand{\diff}{\mathrm{d}}

\begin{document}
\title{On transverse momentum broadening in real-time lattice simulations of the glasma and in the weak-field limit}

\author{\firstname{Andreas} \lastname{Ipp}\inst{1}\fnsep\thanks{\email{ipp@hep.itp.tuwien.ac.at}} \and
        \firstname{David I.} \lastname{M\"uller}\inst{1}\fnsep\thanks{\email{dmueller@hep.itp.tuwien.ac.at}} \and
        \firstname{Daniel} \lastname{Schuh}\inst{1,2}\fnsep\thanks{\email{schuh@hep.itp.tuwien.ac.at}}
}

\institute{Institute for Theoretical Physics, TU Wien,  \\
	Wiedner Hauptstr. 8-10, 1040 Vienna, Austria \and Speaker and corresponding author}

\abstract{In these proceedings, we report on our numerical lattice simulations of partons traversing the boost-invariant, non-perturbative glasma as created at the early stages of collisions at RHIC and LHC. Since these highly energetic partons are produced from hard scatterings during heavy-ion collisions, they are already affected by the first stage of the medium's time evolution, the glasma, which is the pre-equilibrium precursor state of the quark-gluon plasma. We find that partons quickly accumulate transverse momentum up to the saturation momentum during the glasma stage. Moreover, we observe an interesting anisotropy in transverse momentum broadening of partons with larger broadening in the rapidity than in the azimuthal direction. Its origin can be related to correlations among the longitudinal color-electric and color-magnetic flux tubes in the initial state of the glasma. We compare these observations to the semi-analytic results obtained by a weak-field approximation, where we also find such an anisotropy in a parton's transverse momentum broadening.}

\maketitle

\section{Introduction} \label{sec:introduction}
The quark-gluon plasma (QGP) is an extensively researched state of matter that is created after a heavy-ion collision. It is effectively described by viscous hydrodynamics~\cite{Gale:2013, Romatschke:2017}, which allows for the determination of bulk properties, such as particle multiplicities and flow harmonics~\cite{Schenke:2010, Schenke:2011, Schenke:2012, Gale:2012, Niemi:2015}, without exact information about the initial conditions of the hydrodynamic evolution because of the existence of a hydrodynamic attractor~\cite{Berges:2013}. However, there are also probes that carry information throughout the whole evolution, and, hence, they are influenced by the other stages of matter that are created after a heavy-ion collision. Some of these probes do not interact strongly, such as photons and dileptons~\cite{Martinez:2007, Ipp:2009, Ipp:2012, Vujanovic:2014}, while others, such as jets~\cite{Mehtar-Tani:2013, Connors:2017, Busza:2018}, do. The seeds of jets (partons) are created by hard scatterings during the collision and are, therefore, already present in the pre-equilibrium precursor state of the QGP, which is the glasma. The glasma is based on an effective theory of high-energy QCD, namely the color-glass condensate (CGC)~\cite{Gelis:2010, Gelis:2012} and consists of expanding chromo-electric and chromo-magnetic flux tubes~\cite{Lappi:2006}. This state is highly anisotropic because said flux tubes are initially longitudinal and, at sufficiently high energies, boost invariant.

In these proceedings, which are based on~\cite{Ipp:2020a, Ipp:2020b}, we discuss the transverse momentum broadening that a parton accumulates during the glasma stage. We do that in a boost-invariant approximation, which causes the system to be effectively 2+1~dimensional, since its rapidity dependence is dropped. The dilute limit~\cite{Kovner:1995}, in which the color fields are considered to be small, allows for a largely analytic calculation of the transverse momentum broadening. This gives rise to an analytic relation of the momentum broadening anisotropy of an ultra-relativistic test parton to the correlations among the flux tubes in the initial state of the glasma. Recent progress in the dilute limit has been made in 3+1 dimensions~\cite{Ipp:2021} as well. We also discuss transverse momentum broadening in a different limit, namely the lattice approximation, which is a real-time lattice field theoretic approach. It permits the treatment of strong fields in the glasma, providing the opportunity to study this more realistic case. The transverse plane is approximated by a square lattice with periodic boundary conditions, and the discretized Yang-Mills equations are solved with finite differences.

These proceedings are organized in the following way: in sec.~\ref{sec:mom_broad_background_field}, we will discuss the theoretical framework of describing momentum broadening in a non-Abelian background field. Then, in sec.~\ref{sec:mom_broad_glasma}, we will specify the background field to be the glasma and examine two approximations in which we can compute said momentum broadening, namely a lattice approximation (subsec.~\ref{subsec:lattice_approx}) and a weak-field approximation (subsec.~\ref{subsec:weak-field_approx}). We will present our results in sec.~\ref{sec:results} and conclude in sec.~\ref{sec:conclusions}.

\section{Momentum broadening in a non-Abelian background field} \label{sec:mom_broad_background_field}
\begin{figure}
    \centering
    \includegraphics[scale=0.16]{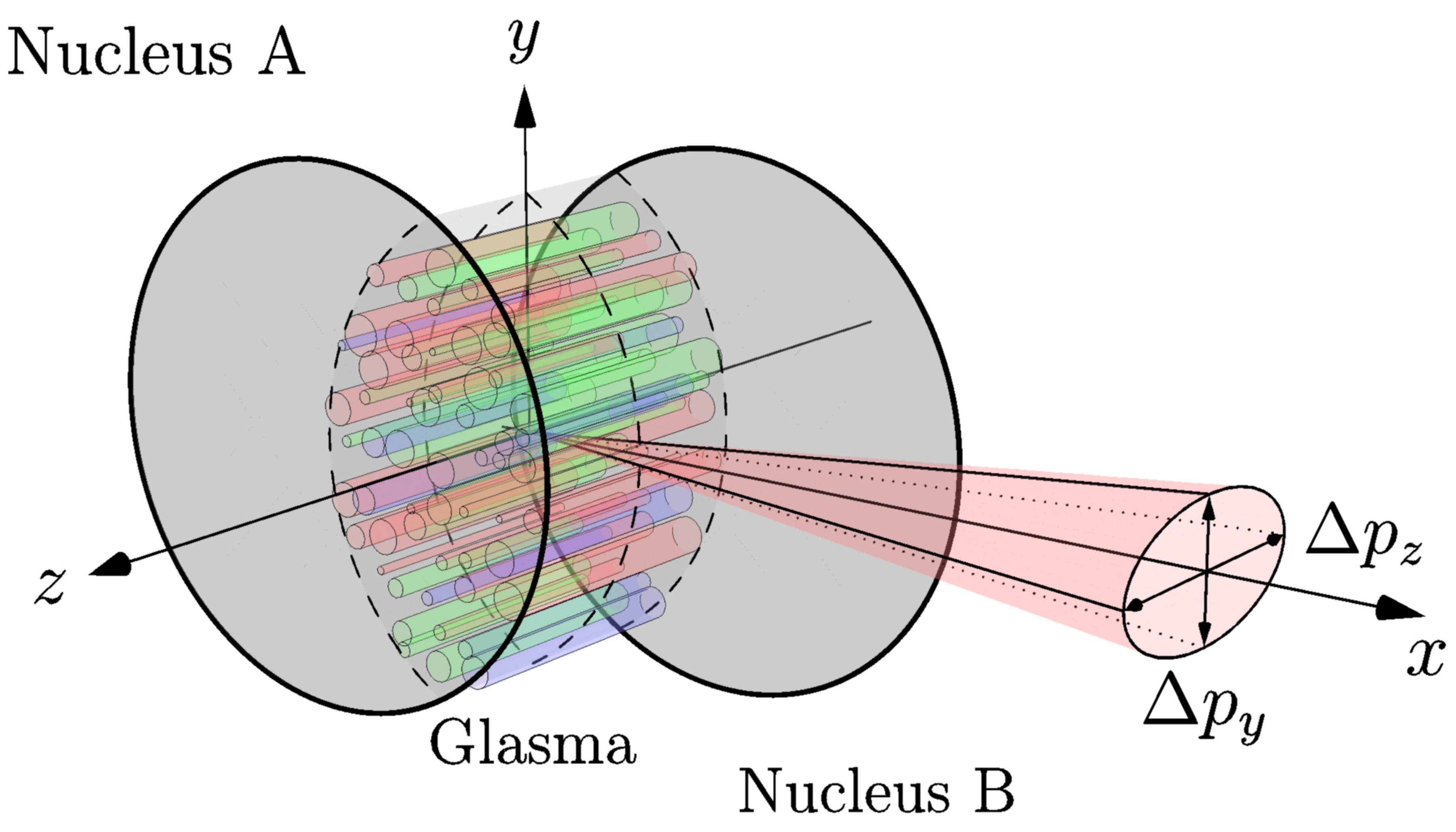}
    \caption{Visualization of a heavy-ion collision, taken from~\cite{Ipp:2020a}. Two colliding nuclei (gray) produce the glasma, which is depicted by colorful cylinders. During the same collision event, a parton (traveling in $x$-direction) is produced and experiences momentum broadening transverse to its movement.}
    \label{fig:momentum_broadening}
\end{figure}

The glasma is produced during the collision of two high-energy nuclei by interactions among their soft partons. We consider an ultra-relativistic parton that is created through hard scatterings during such a collision. We treat it as a test parton and the glasma as a non-Abelian background field, i.e.~back-reactions from the parton on the glasma are neglected, but the glasma exerts a non-Abelian Lorentz force on the parton. Since the parton is ultra-relativistic, its trajectory is considered to be unaffected by this force, but it still accumulates momentum along its path. The concrete geometry is shown in fig.~\ref{fig:momentum_broadening}. The nuclei, depicted in gray, move in positive and negative $z$-direction, respectively, and collide at the origin. The parton travels along the $x$-direction ($u^\mu = (1, 1, 0, 0)^\mu$), the cone indicates the broadening in momentum transverse to the particle trajectory and the glasma is visualized by colorful cylinders.

The methods that we use to obtain an expression for transverse momentum broadening for an ultra-relativistic colored particle are largely based on its definition via lightlike Wilson lines~\cite{Liu:2006, CasalderreySolana:2007, Panero:2013} and the dynamics of classical colored particles~\cite{Majumder:2009, Carrington:2016, Mrowczynski:2017, Ruggieri:2018, Sun:2019}. We review them in~\cite{Ipp:2020a} and also derive a formula for said momentum broadening. For a quark, it reads\vspace{-0.5em}
\begin{equation}
    \ev{p^2_i(t)}_q = \frac{2 g^2}{N_c} \intop_0^t \! \diff t' \intop_0^t \! \diff t'' \, \big< \Tr\big[  U(0, t') F_{i+}(x(t')) U(t', 0) U(0, t'') F_{i+}(x(t'')) U(t'', 0) \big] \big>,\vspace{-0.5em} \label{eq:p_squared_nonabelian}
\end{equation}
where the subscript~$q$ denotes that it holds for a quark, $g$ is the Yang-Mills coupling constant, $N_c$ corresponds to the gauge group~SU$(N_c)$ and~$i \in \{ y, z \}$. The lightlike Wilson line $U(0,t)$ connects the starting position of the particle at its creation at~$t_0 = 0$ with its current position at~$t$, and the field strength tensor is denoted by~$F$. Note that the plus sign in the subscript of the field strength tensor refers to the light-cone coordinates $x^\pm = (t \pm x)/\sqrt{2}$. The transverse momentum broadening for a corresponding gluon is related to eq.~\eqref{eq:p_squared_nonabelian} via Casimir scaling
\begin{equation}
	\ev{p^2_i}_g = \frac{C_A}{C_F} \ev{p^2_i}_q,
\end{equation}
where $C_A$ and $C_F$ are the Casimirs in the adjoint and fundamental representation, respectively.

\section{Momentum broadening in the glasma} \label{sec:mom_broad_glasma}
The glasma is described in the CGC framework. The hard partons of the incoming nuclei~A and~B are described by the color charge densities~$\rho_{(\A, \B)}$. Their color currents in covariant gauge $\p_\mu A^\mu = 0$ are given by $J^\mu_{(\A,\B)} = \delta^\mu_\pm \rho_{(\A,\B)}(x^\mp, \mathbf x)$, with the Kronecker delta~$\delta^\mu_\pm$. The light-cone coordinates are defined as $x^\pm = (t \pm z) / \sqrt{2}$, i.e.~this definition differs by the one used in eq.~\eqref{eq:p_squared_nonabelian} by the replacement $x \leftrightarrow z$. This difference is connected to the fact that in the glasma, the transverse coordinates are viewed as transverse to the beam axis ($z$-axis), while in the context of momentum broadening, the transverse plane is considered to be transverse to the particle direction ($x$-axis).

The charge densities~$\rho_{(\A,\B)}$ are described by the McLerran-Venugopalan model~\cite{McLerran:1994a, McLerran:1994b}. In this model, the nuclei are approximated by infinitely thin color sheets $\rho_{(\A,\B)}(x^\mp, \mathbf x) = \delta(x^\mp)\rho_{(\A,\B)}(\mathbf x)$ that are infinitely big in the transverse directions. Furthermore, the color charge density~$\rho$ is viewed as a random field. Its distribution is given by a Gaussian probability functional~$W[\rho]$, which is specified by the charge density correlator
\begin{equation}
    \ev{\rho^a_{(\A, \B )}(x) \rho^b_{( \A, \B )}(y)} = (g \mu)^{2} \delta^{ab} \delta(x^{\mp} - y^{\mp}) \delta(x^{\mp}) \delta^{(2)}  (\mathbf x - \mathbf y),\vspace{-0.25em}
\end{equation}
with the model parameter~$\mu$ that determines the saturation momentum~$Q_s \propto g^2 \mu$. Expectation values of observables are obtained by functional integration over~$\rho$, weighted by~$W[\rho]$  \vspace{-0.25em}
\begin{equation}
    \ev{\dots} = \int \! \mathcal{D} \rho \, W[\rho] \left( \, \dots \right).\vspace{-0.25em}
\end{equation}
The high-momentum (hard) partons act as sources for the low-momentum (soft) partons. The glasma is obtained from solving the classical Yang-Mills equations in the forward light cone of the collision, which we describe in Milne coordinates. These consist of proper time \mbox{$\tau = \sqrt{2 x^+ x^-}$}, spacetime rapidity~$\eta = \ln (2 x^+ x^-)/2$ and the transverse coordinates~$x$ and~$y$. In the boost-invariant approximation, the Yang-Mills equations are source free and physical observables do not depend on~$\eta$. The glasma initial conditions~\cite{Kovner:1995} in temporal gauge $A^\tau = 0$ read
\begin{equation}
    A^i(0, \vv x) = \alpha^i_\A(\vv x) + \alpha^i_\B(\vv x), \mspace{8mu}
    A^\eta(0, \vv x) = \frac{ig}{2} \left[ \alpha^i_\A(\vv x), \alpha^i_\B(\vv x) \right], \mspace{8mu}
    \p_\tau A^i(0, \vv x) = 0, \mspace{8mu}
    \p_\tau A^\eta(0, \vv x) = 0. \label{eq:ic_glasma1}
\end{equation}
The fields $\alpha_{(\A, \B)}^i$ are given by $\alpha^i_{(\A,\B)}(\vv x) = \frac{1}{ig} V_{(\A,\B)}( \vv x) \p^i V_{(\A,\B)}^\dg( \vv x)$, with the asymptotic Wilson lines
\begin{equation}
    V^\dg_{(\A,\B)}(\vv x) = \lim_{x^\mp \rightarrow \infty} V^\dg_{(\A,\B)}(x^\mp, \vv x) = \lim_{x^\mp \rightarrow \infty} \mathcal{P} \exp{\left(i g \intop^{\infty}_{-\infty} \! \diff x^\mp \, \frac{\rho_{(\A,\B)} (x^\mp , \mathbf x)}{\mathbf \nabla^2 - m^2}\right)} \mspace{2mu},
\end{equation}
where $m$ is an infrared regulator.

Given a solution to the Yang-Mills equations under the glasma initial conditions, we can evaluate the accumulated transverse momentum, given by eq.~\eqref{eq:p_squared_nonabelian}. We obtain
\vspace{-0.25em}
\begin{equation}
	\ev{p^2_i(\tau)}_q = \frac{g^2}{N_c} \intop^\tau_0 \! \diff \tau' \intop^\tau_0 \! \diff \tau'' \, \ev{\mathrm{Tr} \left[ f^i(\tau') f^i(\tau'') \right]}, \vspace{-0.25em}\label{eq:p_suqared}
\end{equation}
with the functions~$f^i$ that represent the color-rotated Lorentz force
\begin{equation}
	f^y(\tau) = U(\tau) \left( E_y(\tau) - B_z(\tau) \right) U^\dg(\tau), \qquad
	f^z(\tau) = U(\tau) \left( E_z(\tau) + B_y(\tau) \right) U^\dg(\tau). \label{eq:f_y_f_z}
\end{equation}
The lightlike Wilson line~$U$ in the fundamental representation along the particle trajectory in temporal gauge is given by\vspace{-0.5em}
\begin{equation}
	U(\tau) = \mathcal{P} \exp{\bigg( - i g \intop_0^\tau \! \diff \tau' \, A_x (\tau') \bigg)}.\vspace{-1.0em}
\end{equation}

\subsection{Lattice approximation} \label{subsec:lattice_approx}
The lattice approximation~\cite{Krasnitz:1998, Lappi:2003} is a real-time lattice gauge theoretic method. We approximate the transverse plane by a square lattice and describe each nucleus by $N_s$ color sheets along the longitudinal direction, which accounts for path ordering~\cite{Fukushima:2007}. The degrees of freedom change to the gauge links in the transverse plane~$U_{x, \hat{i}}$, the rapidity component of the gauge field~$A_{x, \eta}$ and the conjugate momenta~$P^{i}_x$ and $P^{\eta}_x$. The Yang-Mills equations are replaced by a leapfrog scheme for finite time steps, and the glasma initial conditions are discretized. Details of the derivation of the appropriate lattice version of transverse momentum broadening can be found in~\cite{Ipp:2020a}. The result reads
\begin{equation}
    \ev{p^2_{i}(t_n)}_q \approx \frac{g^2 a_T^2}{N_c} \ev{ \Tr \bigg[ \big(\sum^n_{i=0} f^i(t_n) \big)^2 \bigg] }, \vspace{-0.25em}\label{eq:p_suqared_latt}
\end{equation}
where $a_T$ is the transverse lattice spacing, $f^i$ are discretized versions of eq.~\eqref{eq:f_y_f_z} and $t_n$ are the times at which the particle position coincides with a lattice site.

\subsection{Weak-field approximation} \label{subsec:weak-field_approx}
In the weak-field approximation~\cite{Kovner:1995}, the charge density~$\rho$ that describes the nuclei is deemed small. Consequently, path ordering effects do not contribute at the lowest order that is not pure gauge, and the time evolution becomes Abelian at this (fourth) order~\cite{Fujii:2008}. The solution of the color-electromagnetic fields can be written as
\begin{IEEEeqnarray}{rClrCl}
    E_z(\tau, \vv x) &=& \! \int \! \frac{\diff^2 \vv k}{(2 \pi)^2} \, J_0(k \tau) \tilde{E}_z(0, \vv k) e^{i \vv k \cdot \vv{x}},
    & B_z(\tau, \vv x) &=& \! \int \! \frac{\diff^2 \vv k}{(2 \pi)^2} \, J_0(k \tau) \tilde{B}_z(0, \vv k) e^{i \vv k \cdot \vv{x}}, \\
    E_y(\tau, \vv x) &=& \mspace{2mu} - \mspace{-5mu} \int \! \frac{\diff^2 \vv k}{(2 \pi)^2} \, \frac{i k_x}{k} J_1(k \tau) \tilde{B}_z(0, \vv k) e^{i \vv k \cdot \vv{x}}, \mspace{8mu}
    & B_y(\tau, \vv x) &=& \! \int \! \frac{\diff^2 \vv k}{(2 \pi)^2} \, \frac{i k_x}{k} J_1(k \tau) \tilde{E}_z(0, \vv k) e^{i \vv k \cdot \vv{x}}, \IEEEeqnarraynumspace
\end{IEEEeqnarray}
with the Bessel functions~$J_n$ of the first kind, the norm~$k$ of the vector~$\vv k$ and the Fourier components of the initial field strengths~$\tilde{E}_z(0, \vv k)$ and~$\tilde{B}_z(0, \vv k)$. As can be seen from the order of the Bessel function,\footnote{The Bessel function $J_0$ is positive at zero, while the Bessel function $J_1$ vanishes, i.e.~$J_0(0) = 1$ and $J_1(0) = 0$.} the fields are purely longitudinal with respect to the beam axis at~\mbox{$\tau = 0$}. The functions~$f^i$, given by eq.~\eqref{eq:f_y_f_z}, turn into \vspace{-0.5em}
\begin{equation}
    f^y(\tau) = \mspace{2mu} - \mspace{-5mu} \int \! \frac{\diff^2 \vv k}{(2 \pi)^2} \, \gamma(\tau, \vv k) \tilde{B}_z(0, \vv k), \qquad
    f^z(\tau) = \! \int \! \frac{\diff^2 \vv k}{(2 \pi)^2} \, \gamma(\tau, \vv k) \tilde{E}_z(0, \vv k), \label{eq:f_y_f_z_weak}
\end{equation}
with \vspace{-0.5em}
\begin{equation}
    \gamma(\tau, \vv k) = \left( \frac{i k_x}{k} J_1(k \tau) + J_0(k \tau) \right) e^{i k_x \tau}.
\end{equation}
The Wilson line~$U$ does not contribute because it is approximated by a unit matrix at lowest order. The function~$\gamma$ appears in both components of $f^i$, which means that the time dependent part of their integrands is identical. Inserting eq.~\eqref{eq:f_y_f_z_weak} into eq.~\eqref{eq:p_suqared} and evaluating the trace leads to \vspace{-0.5em}
\begin{equation}
    \ev{p^2_{(y, z)}(\tau)}_q = \int \! \frac{\diff^2 \mathbf k}{(2 \pi)^2} \, g(\tau, \mathbf k) \, c_{(B, E)}(k, m), \quad
    g(\tau, \mathbf k) = \frac{N_c^2 - 1}{2} g^8 \mu^4 \left| \intop^\tau_0 \! \diff \tau' \, \gamma(\tau, \vv k) \right|^2 \mspace{-7mu}, \vspace{-0.5em} \label{eq:p_squared_weak}
\end{equation}
with the infrared regulator~$m$. Since $\gamma$ shows up in both components of~$f^i$, the function~$g$ appears in both components of the transverse momentum broadening. They only differ by the initial time correlators~$c_B$ and~$c_E$
\begin{equation}
    c_B(k, m) \mspace{-2mu} = \mspace{-10mu} \int \mspace{-9mu} \frac{\diff^2 \mathbf p}{(2 \pi)^2} \mspace{-1mu} \frac{(\mathbf p \times \mathbf k)^2}{(p^2 \mspace{-4mu} + \mspace{-3mu} m^2)^2 (|\mathbf k \mspace{-1mu} - \mspace{-1mu} \mathbf p|^2 \mspace{-4mu} + \mspace{-3mu} m^2)^2}, \mspace{7mu}
    c_E(k, m) \mspace{-2mu} = \mspace{-10mu} \int \mspace{-9mu} \frac{\diff^2 \mathbf p}{(2 \pi)^2} \mspace{-1mu} \frac{(\mathbf p \cdot (\mathbf k - \mathbf p))^2 }{(p^2 \mspace{-4mu} + \mspace{-3mu} m^2)^2 (|\mathbf k \mspace{-1mu} - \mspace{-1mu} \mathbf p|^2 \mspace{-4mu} + \mspace{-3mu} m^2)^2}.
\end{equation}
This lets us link an anisotropy \mbox{$\ev{p^2_z} / \ev{p^2_y} \neq 1$} in the transverse momentum broadening of an ultra-relativistic test parton directly to the difference between the color-electric and color-magnetic correlators at the time of the collision.

\section{Results and discussion} \label{sec:results}
\begin{figure}
	\centering
	\subfigure[Transverse momentum broadening in the dilute glasma. \label{subfig:mom_broad_weak_field_mom_broad}]{\includegraphics[height=3cm]{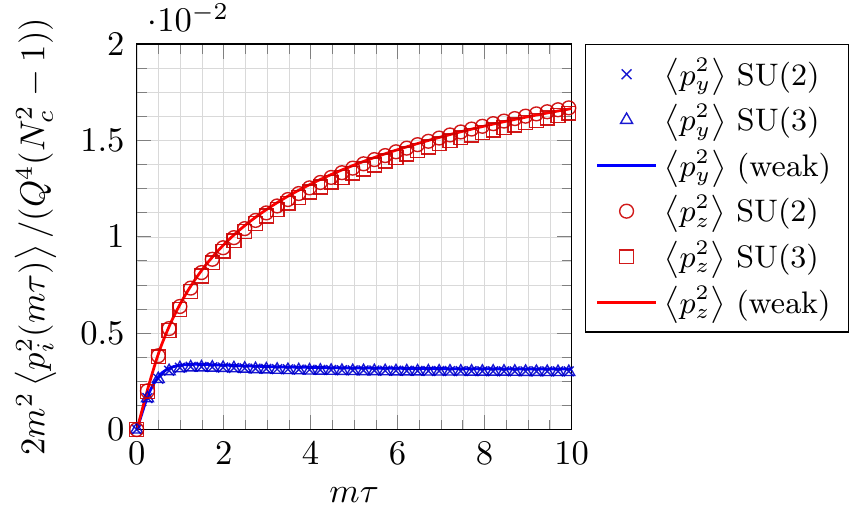}}
	\quad
	\subfigure[Initial correlators of the dilute glasma in momentum space. \label{subfig:mom_broad_weak_field_init_corr_mom_space}]{\includegraphics[height=3cm]{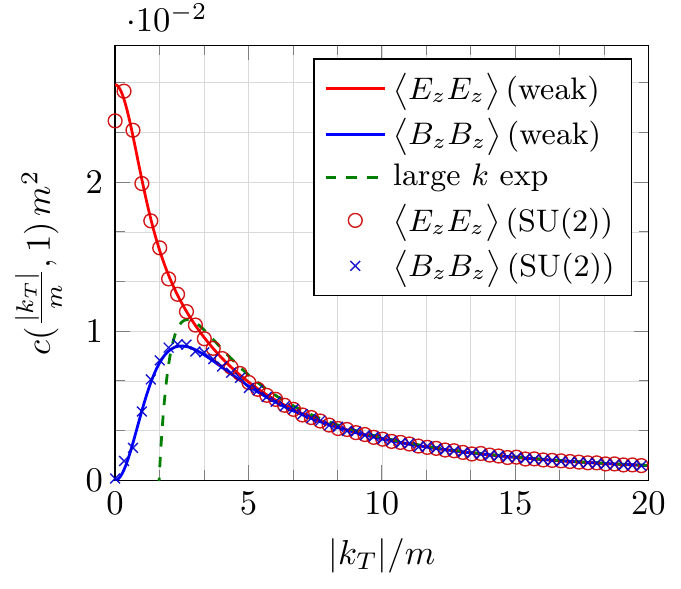}}
	\quad
	\subfigure[Initial correlators of the dilute glasma in coordinate space. \label{subfig:mom_broad_weak_field_init_corr_ccordinate_space} ]{\includegraphics[height=3cm]{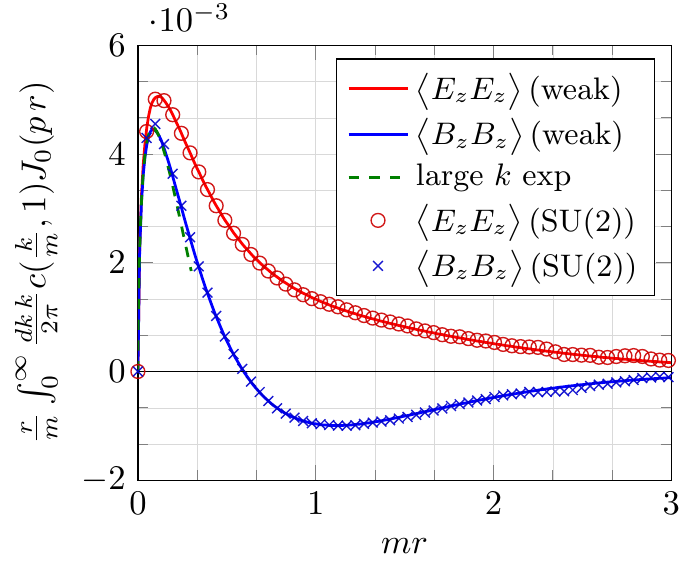}}
	\caption{Transverse momentum broadening and initial field correlators of an ultra-relativistic test parton in the dilute glasma, taken from~\cite{Ipp:2020a}. The lines depict the results of our semi-analytical weak-field approach, and the symbols correspond to our real-time lattice simulations, with SU$(2)$ and SU$(3)$ as gauge groups. Equation~\eqref{eq:p_squared_weak} provides the relationship between (a) and (b, c).}
\end{figure}

Before presenting the results for the physically more relevant lattice simulations that treat strong color fields, we discuss the results of the weak-field limit, in which we have analytic formulas that allow for a better understanding of the mechanisms that lead to momentum broadening. Figure~\ref{subfig:mom_broad_weak_field_mom_broad} displays the accumulated transverse momentum broadening in the dilute glasma. It compares the lattice simulations to the weak-field calculations and shows that they agree for a dilute glasma. Furthermore, it illustrates that the two components of transverse momentum broadening behave differently, which manifests itself by an anisotropy in momentum broadening \mbox{$\ev{p^2_z} / \ev{p^2_y} \neq 1$}. In eq.~\eqref{eq:p_squared_weak}, we linked such an anisotropy to the difference between the color-electric and color-magnetic correlators during the collision, and Figs.~\ref{subfig:mom_broad_weak_field_init_corr_mom_space} and~\ref{subfig:mom_broad_weak_field_init_corr_ccordinate_space} show these correlators in momentum space and in coordinate space, respectively. They agree for large momenta $k \gg m$, but they highly differ from one another in the infrared regime $k < m$. While the electric correlator stays positive as the momentum tends toward zero, the magnetic correlator vanishes \vspace{-0.5em}
\begin{equation}
    \lim_{k \rightarrow 0} \, c_E(k, m) = \frac{1}{12 \pi^2 m^2} > 0, \qquad \lim_{k \rightarrow 0} \, c_B(k, m) = 0.
\end{equation}
In coordinate space, the difference between the two correlators manifests itself in that the electric correlator is positive everywhere, while the magnetic correlator exhibits a region of anti-correlation around $mr \approx 1$. This anti-correlation is responsible for the flattening of~$\ev{p^2_y}$. Note that it is the infrared modes of the correlators that differ and are, therefore, responsible for the anisotropy in transverse momentum broadening and that observables that are less sensitive to these modes show a smaller discrimination between electric and magnetic flux tubes. The energy density of electric and magnetic flux tubes, for example, is very similar because there, the infrared modes are suppressed.

\begin{figure}
    \centering
    \subfigure[Momentum broadening in the dense glasma at early times. \label{subfig:mom_broad_lattice_early_time}]{\includegraphics[height=3cm]{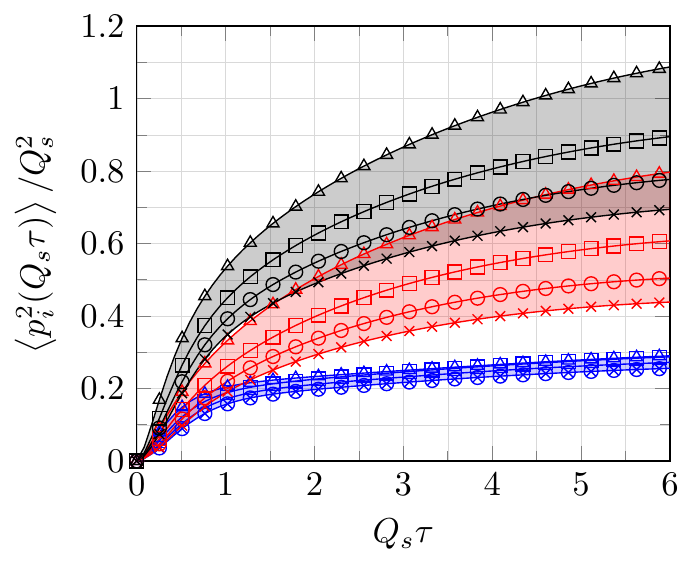}}
    \qquad
    \subfigure[Momentum broadening in the dense glasma up to later times. \label{subfig:mom_broad_lattice_late_time}]{\includegraphics[height=3cm]{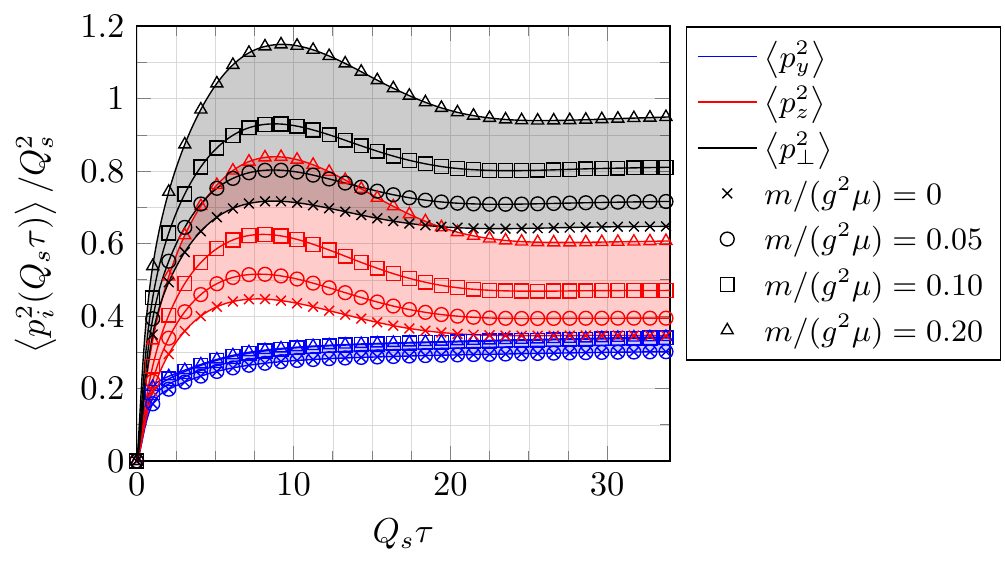}} \\
    \subfigure[Initial correlators of the dense glasma in momentum space. \label{subfig:mom_broad_lattice_init_corr_mom_space}]{\includegraphics[height=3cm]{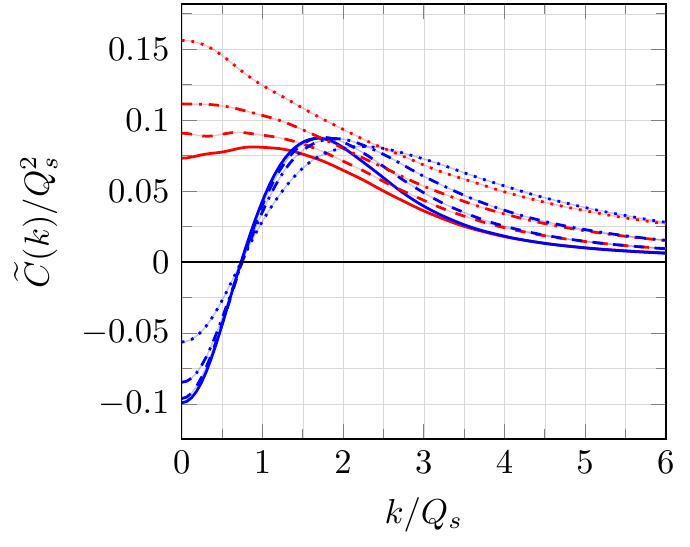}}
    \qquad
    \subfigure[Initial correlators of the dense glasma in coordinate space. \label{subfig:mom_broad_lattice_init_corr_coordinate_space}]{\includegraphics[height=3cm]{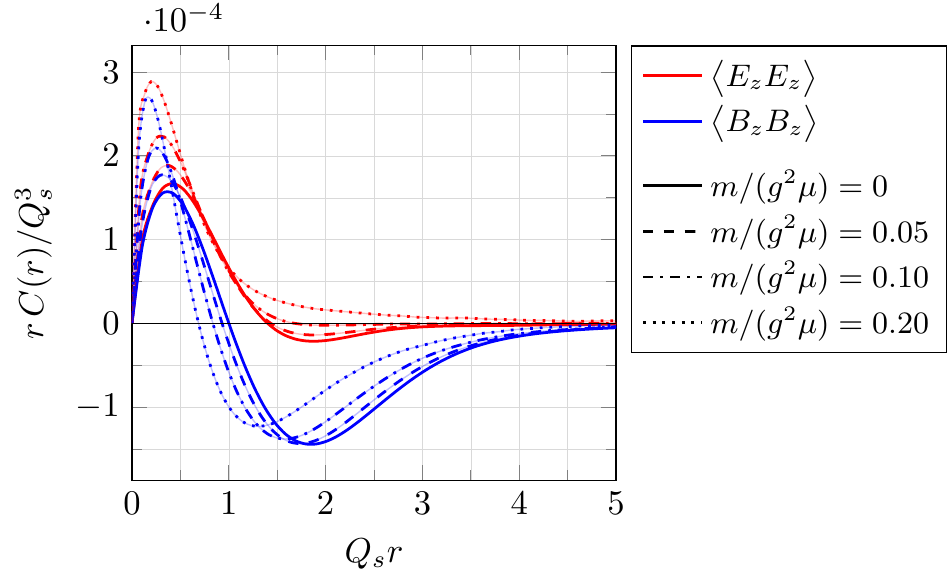}}
    \caption{Transverse momentum broadening and initial field correlators of an ultra-relativistic test parton in the dense glasma from SU$(3)$ real-time lattice simulations, taken from~\cite{Ipp:2020a}. The different symbols correspond to various densities of the glasma, determined by the ratio $m / (g^2 \mu)$, and the bands indicate the values that momentum broadening can take for the chosen ratios. The initial color-electric correlator in coordinate space is defined as $C_E(|\vv x - \vv y|) = \ev{\Tr \left[ E_z(\vv x) U_{\vv x \rightarrow \vv y} E_z(\vv y) U_{\vv y \rightarrow \vv x}  \right]}$, where the Wilson line~$U_{\vv x \rightarrow \vv y}$ connects the lattice points $\vv x$ and $\vv y$. The color-magnetic correlator is defined analogously ($E \rightarrow B$) and $\widetilde{C}(k)$ are the Hankel transforms of order zero. \vspace{-2em}}
\end{figure}

Figures~\ref{subfig:mom_broad_lattice_early_time} and~\ref{subfig:mom_broad_lattice_late_time} depict the momentum broadening for strong fields $m / (g^2 \mu) \ll 1$. The behavior at early times looks qualitatively similar to the dilute glasma, while $\ev{p^2_z}$ differs at later times in that it starts to decrease just below $Q_s \tau = 10$, as does the total transverse momentum broadening $\ev{p^2_\perp} = \ev{p^2_y} + \ev{p^2_z}$. The $y$-component of transverse momentum broadening depends only weakly on the density of the glasma, whereas the numerical values of the $z$-component highly differ for varying $m / (g^2 \mu)$, so much so that the anisotropy vanishes at very late times for a very strong glasma. Note, however, that at such late times ($Q_s \tau \gtrsim 10$), the glasma is no longer a valid description for the state of matter that is produced in heavy-ion collisions. Figures~\ref{subfig:mom_broad_lattice_init_corr_mom_space} and~\ref{subfig:mom_broad_lattice_init_corr_coordinate_space} show the initial correlators for strong fields in momentum and coordinate space, respectively. While each of them individually depends on the glasma density numerically, there is also a qualitative difference between them. In momentum space, they differ at low momenta, and the color-electric correlator tends towards a positive value as the momentum approaches zero. The color-magnetic correlator admits a negative value for vanishing momentum, which becomes closer to zero as the glasma becomes more dilute, as one would expect from the weak-field results, where it vanishes for $k = 0$. In coordinate space, the color-magnetic correlator looks qualitatively similar to the corresponding weak-field correlator, with increasing similarity for rising $m / (g^2 \mu)$. The color-electric correlator, however, exhibits a negative region for glasmas denser than $m / (g^2 \mu) \lesssim 0.1$. This is in contrast to the dilute glasma. Note that despite some similarities in the results of transverse momentum broadening and the initial correlators between the dilute and the dense glasma, we do not have an analytic relation that lets us link the momentum broadening anisotropy to the initial time correlators for a dense glasma. Nevertheless, we expect a relation akin to eq.~\eqref{eq:p_squared_weak} due to these similarities. \vspace{-0.5em}

\section{Conclusions and outlook} \label{sec:conclusions}
In these proceedings, we discussed transverse momentum broadening of an ultra-relativistic test parton in the glasma, which is the precursor state of the QGP. After revisiting the formalism that provides the basis for our calculations, we presented two approximations with which we were able to compute said momentum broadening and analyzed their similarities and differences. In the weak-field approximation, we found a formula that provides a link between the anisotropy in transverse momentum broadening and the difference in the low momentum modes of the color-electric and color-magnetic correlators at the time of the collision. For realistic fields, we found somewhat similar results in the relevant time scales, which suggests that the weak-field limit can be seen as an interesting approximation in which the underlying mechanisms can be studied analytically.

In order to extend this work, one could use more realistic initial conditions for the glasma or relax the ultra-relativistic test particle approximation. The former contains the treatment of the nuclei as having finite thickness~\cite{Ipp:2021}, which implies a rapidity dependence of the system, and finite spatial extent in the transverse direction, which allows to study off-central collisions. The latter leads to a deflection of the parton from its straight trajectory and to the inclusion of its back-reaction on the glasma. It would also be interesting to study other observables, such as the energy loss of a test parton traversing the glasma, in order to better understand the early-time behavior of partons that create jets at later stages.

\vspace{0.5em}

\begin{acknowledgement}
This work has been supported by the Austrian Science Fund FWF No.~P32446-N27 and No.~P28352. 
The Titan\,V GPU used for this research was donated by the NVIDIA Corporation.
\end{acknowledgement}

\bibliography{references.bib}

\end{document}